\documentclass{article}

\usepackage[margin=1in]{geometry}

\usepackage{graphicx} 
\usepackage{amsmath}
\usepackage{amssymb}
\usepackage{dsfont}
\usepackage{appendix}

\usepackage{color}

\newcommand{\bra}{\langle}
\newcommand{\ket}{\rangle}

\newcommand{\tl}[1]{\tilde{#1}}

\newcommand{\pdr}{\partial}

\newcommand{\beq}{\begin{equation}}
\newcommand{\eeq}{\end{equation}}
\newcommand{\beqs}{\begin{eqnarray}}
\newcommand{\eeqs}{\end{eqnarray}}

\newcommand{\ov}[1]{\frac{1}{#1}}
\newcommand{\fr}[2]{\frac{#1}{#2}}

\def\al{\alpha}

\def\del{\delta}

\def\th{\theta}	
\def\Th{\Theta}

\def\Ms{{\rm Ms}}
\def\Mc{{\rm Mc}}
\def\s{{\rm s}}
\def\tr{{\rm tr}}

\title{An Open Quantum System of Coupled Rotors.}
\date{}
\begin{document}
\maketitle
	\baselineskip=24pt
\begin{center}
\large
\author{V V Sreedhar\footnote{sreedhar@cmi.ac.in}}
and 
\author{Ankit Yadav\footnote{ankit@cmi.ac.in}}
\end{center}

        \centerline{\it Chennai Mathematical Institute,  SIPCOT IT Park, Siruseri, Chennai, 603103 India}

\bigskip
\begin{abstract}
A quantum mechanical system of two coupled rotors (particles constrained to 
move on a circle) is studied from an open quantum systems point of view. 
One of the rotors is integrated out and the reduced density operator of 
the other rotor is studied. It's eigenvalues are worked out explicitly using
the properties of Mathieu functions, and the von Neumann entropy, which is a 
standard measure of entanglement, is computed in terms of the Fourier 
coefficients defining the Mathieu functions. Furthermore, upon introducing a time-periodic delta kick and making one of the rotors much heavier than the other, the two-rotor system can be interpreted as a system-bath model, allowing us to introduce a series of approximations to derive a 
master equation of the Lindblad type describing the time-evolution of the 
reduced density operator.   
\end{abstract}

\section{Introduction}

Open quantum systems \cite{breur} arise naturally when one considers the 
effect of the rest of the universe on the system of interest, with the former 
often modelled by a bath of simple harmonic oscillators \cite{caldiera-leggett}.
Their study is especially important in condensed matter physics to understand 
issues like dissipation and diffusion \cite{ballentine}. Besides, entanglement 
which plays a key role in quantum computation \cite{chuang}, and decoherence 
which underpins the quantum-classical transition \cite{zurek}, are all 
naturally studied in the context of open quantum systems. For obvious reasons, 
most papers in the field deal with one or more qubits coupled to a bath of 
oscillators. Some work has also been done by considering quantum rotors -- 
particles restricted to move on a circle \cite{puebla}. When subjected to a 
(quasi-)periodic delta-function kick in time, such systems are known to 
exhibit chaos at the classical level \cite{arul, santhanam}, which has 
interesting ramifications for dynamical localization at the quantum 
level \cite{delande}. Systems subject to periodic temporal perturbations are 
called Floquet systems \cite {reichl} and are an active field of research \cite{bukov, mori, buch}. 

In the present paper, we examine some of the above aspects in the context of 
two coupled quantum rotors, both with and without kicks. The contents of this 
paper are arranged as follows: In the next section, we study the two-rotor 
problem. This entails studying the entanglement between two particles 
constrained to move on a circle interacting via a cosine potential between 
their angle coordinates. The center of mass of the system decouples and is 
easily solved in terms of the eigenfunctions of a free particle on a circle. 
The relative coordinate, however, is given by a more complicated equation, 
that which appears in the solution of pendulum, namely the Mathieu equation. 
By integrating out one of the particles, we derive the reduced density matrix 
associated with the other particle, and thence, the von Neumann entropy is 
obtained in terms of the Fourier coefficients that define the relevant Mathieu 
functions.  

In the next section, we modify the potential between the rotors to include a 
delta-function time periodic kick. The addition of the kick makes the classical
problem chaotic, a fact used in \cite{benenti} to mimic the effect of the bath through a kicked rotor. Here, we consider one particle as the bath and other as the system by making the system Hilbert space much smaller than the bath Hilbert space. We demonstrate that the dynamics becomes Markovian by constructing a CP-divisible map  for the system \cite{rivas}. In section \ref{GSKL},  we show that after making a series of physically well-motivated
approximations, we can obtain the well-known 
Gorini-Kossakowski-Sudarshan-Lindblad equation \cite{lindblad}, for the 
reduced density matrix of the system rotor. 

In the last section, we conclude and present an outlook.

\section{The Two-Rotor System}

The Hamiltonian of the two-rotor system we consider is
    \begin{equation}
        H = -\frac{\hbar^2}{2 m r^2} (\partial_{\theta_1}^2 + \partial_{\theta_2}^2) + g (1 - \cos(\theta_1 - \theta_2)).
    \end{equation}
The configuration space of this system is $S^1 \times S^1$, implying 
periodic boundary conditions on the angle variables to ensure 
single-valuedness of the wavefunction $\psi(\th_1, \th_2)$. The 
radius $r$ is taken to be a constant.
    \begin{equation}
        \psi(\th_1, \th_2) = \psi(\th_1+ 2\pi, \th_2) = \psi(\th_1, \th_2+ 2\pi) = \psi(\th_1+ 2\pi, \th_2 + 2\pi).
    \end{equation}
We can decouple the Hamiltonian by going to the centre of mass $\Theta = 
(\theta_1+\theta_2)/2$ and the relative $\theta = (\theta_1-\theta_2)/2$ 
coordinates. The Hamiltonian in these coordinates (and their conjugate momenta) is
    \begin{equation}
        H = H_\Th + H_\th = \frac{-\hbar^2}{4 m r^2} \pdr_{\Th}^2 + \left( \frac{-\hbar^2}{4 m r^2}\pdr_{\th}^2 + g (1 - \cos 2 \th) \right).
    \end{equation}
Separating variables, the time-independent Schr\"odinger equation with the 
above Hamiltonian breaks up into the following equations:
    \begin{equation}
        \frac{-\hbar^2}{4 m r^2} \pdr_{\Th}^2 \phi_1^m = E_1^m \phi_1^m \quad \text{ and } \quad \frac{-\hbar^2}{4 m r^2}\pdr_{\th}^2 \phi_2^n + g (1 - \cos 2 \th)\phi_2^n = E_2^n \phi_2^n.
        \label{decoupled-eqn}
    \end{equation}
From here onwards we will work in units where $\hbar = m = 2r = 1$. The Schrodinger equation for $\Th$ is that of a free particle on a ring, while that of $\th$ is a Matheiu equation $y'' + (a + q \cos 2x) y = 0$ for pendulum, with parameters $a = E_2 - g$ and $q = g$. The boundary conditions for the wavefunction in these new variables are
    \begin{equation}
        \phi_1(\Th) = \phi_1(\Th + 2\pi), \; \phi_2(\th) = \phi(\th + 2 \pi) \quad \text{or} \quad \phi_1(\Th) = \phi_1(\Th + \pi), \; \phi_2(\th) = \phi(\th + \pi).
    \end{equation}
Using these boundary conditions we obtain the solutions for both the Schr\"odinger equations
    \begin{equation}
    \begin{split}
        \phi_1^m(\Th) = \ov{\sqrt{2 \pi}} e^{i m \Th} \quad& \text{where $m = \sqrt{E_1^m} \in \mathbb{Z}$ and} \\
        \phi_2^n(\th) = \ov{\sqrt{\pi}}\Mc/\s_{2n}(\th) \quad \text{if $m$ is even}, &\quad \phi_2^n(\th) = \ov{\sqrt{\pi}}\Mc/\s_{2n+1}(\th) \quad \text{if $m$ is odd.}
    \end{split}
    \end{equation}
Here $\Mc_k$ is a Mathieu cosine and $\Ms_k$ is a Mathieu sine function of order $k \in \mathbb{N}$ \cite{matthew-walker}. These functions are $\pi$-periodic when $k$ is even and $2 \pi$-periodic otherwise. Now we will consider only even $m$ and $\pi$-periodic Mathieu functions.

\begin{figure}
    \centering
    \includegraphics[width= 0.7\textwidth]{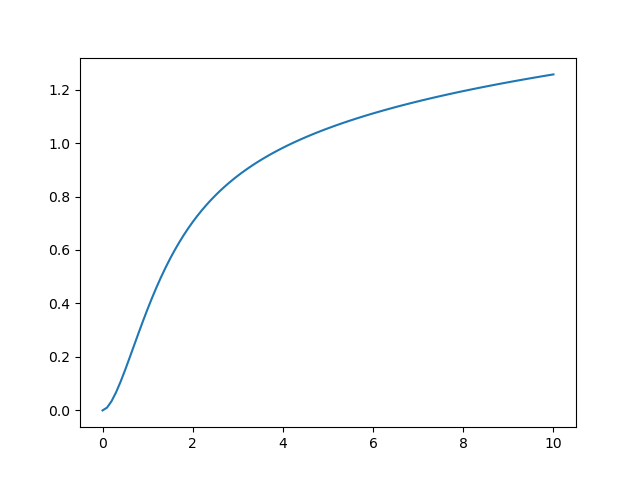}
    \caption{Entanglement entropy $S$ vs the coupling strength $g$ 
for the ground state of the two-rotor system.}
    \label{ent-entropy}
\end{figure}

The ground state of this system is
    \begin{equation}
        \psi^0(\Th, \th) = \phi_1^0(\Th) \phi_2^0(\th) = \ov{\sqrt2 \pi} \Mc_0(\th)
        = \ov{\sqrt2 \pi} \sum_{k = 0}^{\infty} A_{2k} \cos 2k\th = \ov{\sqrt2 \pi} \sum_{k = 0}^{\infty} A_{2k} \cos k(\th_1 - \th_2), 
    \end{equation}
where we have used the standard Fourier expansion of the Mathieu functions in
terms of trigonometric functions. The next step involves
constructing the density matrix of the system and tracing out the $\th_2$ 
variable to get the reduced density matrix in terms of $\th_1$ variable
    \begin{equation}
    \begin{split}
         \rho_1(\th_1, \th_1') = \int_0^{2 \pi} \rho(\th_1, \th_2, \th_1', \th_2) d \th_2 = \int_0^{2 \pi} \psi^0(\th_1, \th_2) \bar{\psi^0}(\th_1', \th_2) d \th_2 \\
         = \ov{ \pi} \left(|A_0|^2 + \ov{4} \sum_{k=1}^\infty |A_{2k}|^2 (e^{ik(\th_1 - \th_1')} + e^{-ik(\th_1 - \th_1')})\right)
    \end{split}
    \end{equation}
To obtain the eigenvalues of the reduced density matrix, we solve the 
eigenvalue equation 
    \begin{equation}
        \int_0^{2\pi} \rho_1(\th_1, \th_1') f_n (\th_1') d\th_1' = p_n f_n(\th_1)
    \end{equation}
The eigenfunctions of the reduced density matrix are same as the energy eigenfunctions for free particle on a ring, $1/\sqrt{2\pi}$ $ exp(ik \th_1')$, where $k \in \mathbb{Z}$.  The eigenvalues of the reduced density matrix are $p_0 = 2 |A_0|^2$ and $p_k = p_{-k} = |A_{2k}|^2/2$. It follows that the entanglement entropy is
    \begin{equation}
        S = - \sum_{n} p_n \ln{p_n} = -2 |A_0|^2 \ln{(2|A_0|^2)} -\sum_{k=1}^\infty |A_{2k}|^2 \ln{(|A_{2k}|^2/2)}
    \end{equation}
We have done the above calculation for the ground state but it holds for any 
energy eigenstate of total system; just the Fourier coefficients will be 
different. For $q = 1$, we get the entropy $S = 0.38$ and the energy of the 
system is $E = E_2 = 0.545$ in appropriate units. The entanglement entropy $S$
is plotted as a function of the coupling strength $g$, as well as energy ( as is clear from the discussion following Eq. \ref{decoupled-eqn}), 
and is observed to increase monotonically (Fig.~\ref{ent-entropy}), as expected. 

\section{The System-Bath model}

We now consider two free particles on a unit circle coupled by a kicked potential. The hamiltonian of the system is
\beq
\hat H = \hat H_S + \hat H_B + \hat H_I,
\eeq
where $ \hat H_{S,B} = \fr{\hat p_{S,B}^2}{2m_{S,B} \, r^2} = \fr{\hat p_{S,B}^2}{2m_{S,B}}$, describe the free rotors and
\beq
\hat H_{I} = g \cos( \hat \th_S - \hat \th_B) \sum_n \del(t - n \tau)
\eeq
the interaction between two rotors. The interaction acts periodically, only for an instance, after a regular interval of time $\tau$. Here $g$ denotes the kick strength between the rotors. In a reference frame fixed to the 
system rotor, the other rotor which is kicked periodically produces a classically chaotic 
environment which mimics the effects of a many-body bath, as emphasised by 
Rossini, Benenti, and Casati \cite {benenti}. Thus, without further ado, we 
consider one of the rotors to be the system and other as bath. The Hilbert space for the total system is $\mathcal{H} = \mathcal{H}_S \otimes \mathcal{H}_B$, with dimensions $dim(\mathcal{H}_S) \times dim(\mathcal{H}_B)$, where $\mathcal{H}_S$ denotes system Hilbert space and $\mathcal{H}_B$ denotes bath Hilbert space.

Due to the explicit time dependence of the total Hamiltonian, time translation invariance is broken. However, because of the $\tau$ periodicity of the kicks, we have a discrete time translation symmetry: $\hat H (t+n \tau) = \hat H(t)$ for $n \in \mathbb{Z}$ and $t \in \mathbb{R}$. So, it becomes convenient to study the dynamics in discrete time steps of $\tau$. The time evolution operator for one kick, called the Floquet operator, is
\beq
\hat U(\tau,0) = e^{-i g \cos (\hat \th_S - \hat \th_B)/ \hbar} e^{-i \hat p_S^2 \tau / (2 m_S \hbar)} e^{-i \hat p_B^2 \tau/(2 m_B \hbar) }.
\eeq
Henceforth, we take $\hbar =1$. Using the discrete time translation invariance, the time evolution for $n$ kicks can be written in terms of the Floquet operator as $\hat U(n \tau,0) = \hat U(\tau,0)^n $.

Initially, {\it i.e.} at $t=0$, we assume that the environment and the system 
have no interaction and the initial density matrix can be represented by a 
separable density matrix $\rho(0) = \rho_S(0) \otimes \rho_B(0)$. The bath is 
chosen to be in some state
\beq
\rho_B = \sum_\nu c_\nu |\nu \ket \bra \nu| \quad \text{where} \quad 0 \leq c_\nu \leq 1.
\label{bath-dens-mat}
\eeq
Here $\sum c_\nu = 1$ and $\{|\nu\ket\}$ furnish an orthonormal basis of the 
bath Hilbert space $\mathcal{H}_B$. For our convenience, we take $|\nu\ket$ 
to be the eigenbasis of the free bath Hamiltonian, $\bra\th| \nu \ket = 
e^{i \nu \th}/\sqrt{2 \pi}$. The unitarily evolved density matrix after 
$n$ kicks is
\beq
\rho(t = n \tau) = U(n \tau,0) \rho_S(0) \otimes \rho_B(0) U(n \tau,0)^\dagger.
\eeq
We can study the dynamics of the system rotor $\rho_S$ in the Kraus Operator 
Sum Representation (OSR)\cite{kraus}
\beq
\rho_S(n \tau)  \equiv \tr_B(\rho(n \tau)) = \int d \th_B \sum_\nu c_\nu \bra \th_B | \hat U(n \tau) | \nu \ket \rho_S(0) \bra \nu | \hat U^\dagger (n \tau) | \th_B \ket = \sum_\nu \int d\th_B K_{\th \nu}(n \tau) \rho_S(0) K^\dagger_{\th \nu}(n \tau),
\label{red-density-matrix}
\eeq
where
\beqs
K_{\th \nu} (n \tau) &=& \sqrt{c_\nu} \sum_{\mu_1} \cdots \sum_{\mu_{n-1}} \idotsint  d\th_1 \cdots d \th_{n-1} \bra \th_b | \hat U(n \tau ,(n-1) \tau) | \mu_{n-1} \ket \bra \mu_{n-1}| \th_{n-1}\ket \bra \th_{n-1}| \cr
&&  \hat U((n-1) \tau, (n-2) \tau) | \mu_{n-2} \ket \bra \mu_{n-2}| \th_{n-2}\ket \bra \th_{n-1}| \cdots
\hat U(2 \tau, \tau)| \mu_1 \ket \bra \mu_1| \th_1\ket \bra \th_1| U( \tau, 0)| \nu \ket
\label{kraus-operaor-n}
\eeqs
are the Kraus operators for the system.
Here, $\bra \mu_j | \th_j \ket = e^{-i \mu_j \th_j}/\sqrt{2 \pi}$ represents a bath state and 
\beq
\bra \th_j | U (j \tau ,(j-1) \tau) | \mu_{j-1}\ket = e^{-i g \cos (\th_j - \hat \th_S)} e^{-i \mu_{j-1}^2 \tau / (2 m_B)} e^{i \mu_{j-1} \th_j} e^{-i \hat p_S^2 \tau /(2 m_S)}.
\eeq

\subsection{Stationary bath assumption}
Up until now both the particles are at same footing. Now, to treat one particle as the bath and other as the system, we take $dim(\mathcal{H}_B) \gg dim(\mathcal{H}_S)$, where the $dim(\mathcal{H}_{B}) \approx m_{B}/ \tau \to \infty$ (in units $\hbar = r=1$). Now, if we put a cut off at the high energies of the bath ($c_\nu = 0$ for $\nu>>1$ in Eq. \ref{bath-dens-mat}), we can take the bath to be stationary $\rho_B (n\tau) = \rho_B(0)$. This assmption is based on the observation that
\beq
\int d\th_1 e^{-i \th_j (\mu_1- \nu)} e^{-i g \cos(\th_1 - \hat \th_S)} = 0,
\label{e:rapid-osc}
\eeq
for $|\mu_1 - \nu| >>1$. Using (\ref{e:rapid-osc}) and the limit $m_B/\tau \to \infty$, we can take
\beqs
&& \int_0^{2 \pi} d\th_1 \sum_{\mu_1 = -\infty}^{\infty} e^{-i \mu_1^2 \tau /(2 m_B)} \fr{e^{i \mu_1 (\th_2 - \th_1 )}}{(2 \pi)^{3/2}} e^{-i ( g \cos(\th_1 - \hat \th_S)- \nu \th_1)} \cr
= && \int_0^{2 \pi} d\th_1 \sum_{\mu_1 = -\infty}^{\infty} \fr{e^{i \mu_1 (\th_2 - \th_1 )}}{(2 \pi)^{3/2}} e^{-i ( g \cos(\th_1 - \hat \th_S) - \nu \th_1)}  \cr
= && \int_0^{2 \pi} d\th_1 \del (\th_2 - \th_1 ) \, e^{-i ( g \cos(\th_1 - \hat \th_S) - \nu \th_1)} / \sqrt{2 \pi}
= e^{-i ( g \cos(\th_2 - \hat \th_S) - \nu \th_2)}/\sqrt{2 \pi}.
\eeqs
Here we have used the fact that for finite $\mu_1$ and $m_B \to \infty$, we can take $e^{-i \mu_1^2 / (2 m_B)} \to 1$.

Using the above simplifications, we can write (\ref{kraus-operaor-n}) as
\beq
K_{\th \nu} (n \tau) = \sqrt{c_\nu} \fr{e^{i \nu \th}}{\sqrt{2 \pi}} \left( e^{-i g \cos(\th - \hat \th_S)} e^{-i \hat p_S^2 \tau /(2 m_S)} \right)^n,
\eeq
which is similar to obtaining the reduced system dynamics interacting with a stationary bath (see Section \ref{markovian-bath}).  
It's easy to verify that $\sum_\al (K^\al)^\dagger K^\al = \mathds{1}$
\beqs
\sum_\nu \int_0^{2 \pi} d\th K_{\th \nu}^\dagger K_{\th \nu} = \sum_\nu c_\nu \int_0^{2 \pi} \fr{d \th}{2 \pi} \left( e^{i \hat p_S^2 \tau /(2 m_S)} e^{i g \cos(\th - \hat \th_S)} \right)^n e^{-i \nu \th} \cr
\times \, e^{i \nu \th} \left( e^{-i g \cos(\th - \hat \th_S)} e^{-i \hat p_S^2 \tau /(2 m_S)} \right)^n = \ov{2\pi} \int_0^{2 \pi} d \th \; \mathds{1} = \mathds{1}
\eeqs
The reduced density matrix for the system (\ref{red-density-matrix}) can therefore be written as
\beq
\rho_S (n \tau) = \ov{2 \pi} \int_0^{2 \pi} d \th \, \left( e^{-i g \cos(\th - \hat \th_S)} e^{-i \hat p^2 \tau /(2 m_S)} \right)^n \rho_S(0) \left( e^{i \hat p^2 \tau /(2 m_S)} e^{i g \cos(\th - \hat \th_S)} \right)^n =\Phi_{n \tau}[\rho_S(0)].
\label{non-markov-sys}
\eeq
Note that the dynamical map $\Phi_t$ is a complete positive trace preserving (CPTP) map as it is derived from the Kraus operator sum representation of the reduced density matrix \cite{kraus}. This ensures that $\rho_s(t)$ is a physical density matrix. Although, we have managed to achieve a compact expression for the reduced density matrix $\rho_\s(t)$ in terms of the initial density matrix $\rho_\s(0)$, finding the explicit form of density matrix at $t$ is still a cumbersome task. The difficulty lies in the interaction terms. If we take 
$g = 0$, the dynamics of the system decouples from the bath and the system 
evolves unitarily in time.
\beq
\rho_S (t= n\tau) = U_S (t;0) \rho(0) U_S^\dagger(t;0) = e^{-i p^2 \tau/(2 m_S)} \rho(0) e^{i p^2 \tau/(2 m_S)} ,
\eeq
and the dynamics is $CP$ divisible \cite{chrusc}. We have to make some 
approximations to make the dynamics CP divisible for the case of non-zero 
interaction. We turn to this issue in the next section.

\subsection{Markovian bath assumption}
\label{markovian-bath}

In this section, we try to make the bath Markovian by using some approximations. First, we make the {\it Born approximation}
\beq
\rho(t) \approx \rho_S (t) \otimes \rho_B (0) = \rho_\s \otimes \rho_{\rm th},
\eeq
where $\rho_{\rm th}$ is the equilibrium  thermal density matrix for bath at inverse temprature $\beta$. This approximation is valid in the regime where (a) the system's Hilbert space is finite-dimensional, while the bath's 
Hilbert space is infinite-dimensional, and (b) the interaction between the 
bath and the system is weak.

Recall that we have already shown in the last section, that the bath can be 
taken to be stationary if we start with the bath density matrix in (or near) the 
ground state . However, for weak kick strength $g$ also the bath can be taken to be stationary. To show that, we expand the interaction term in the time evolution 
operator for one kick to the linear order in $g$
\beq
\hat U(\tau,0) = e^{-i g \cos (\hat \th_B - \hat \th_S)} e^{-i \hat p_S^2 
\tau / (2 m_S)} e^{-i \hat p_B^2 \tau/(2 m_B) } = \left(\mathds{1} - i g 
\cos (\hat \th_S - \hat \th_B)\right) e^{-i \hat p_S^2 \tau / (2 m_S)} 
e^{-i \hat p_B^2 \tau/(2 m_B) }.
\label{time-evol-linear}
\eeq
Additionally, we can simplify the bath density matrix by taking $m_B/\tau \to 
\infty$ and with a cut-off at higher energy levels, $c_n = 1$ for $|n|<N_0$ 
and $c_n = 0$ otherwise, where $N_0 \gg 1$ and $N_0^2 \tau/m_B \ll 1$. In this 
limit, the bath density matrix
\beq
\rho_{\rm th} = \fr{\sum_n c_n e^{-\beta E_n}|n \ket \bra n |} {{\rm tr}(\sum_n c_n e^{-\beta E_n})} \approx \fr{\mathds{1}}{2 N_0}
\label{bath-simp}
\eeq
simplifies to be a constant matrix (to be precise a diagonal matrix with 
entries 0 and 1).

\begin{figure}
    \centering
    \includegraphics[width=0.7\linewidth]{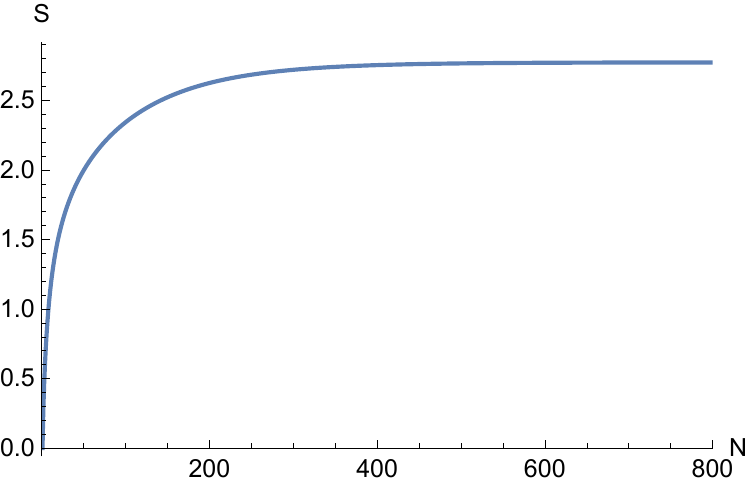}
    \caption{The von-Neumann entropy $S$ vs time $\tau$ plot for kick strength $g=0.1414$ using (\ref{CP-2-rotor-map}), we observe that the entropy increases monotonically. We have started with the system initially in the ground state.}
    \label{fig:CP-map}
\end{figure}

Using these two simplifications we get the bath density matrix as
\beq
\rho_B(\tau) = {\rm tr}_S \left(\hat U(\tau,0) \rho(0) \hat U^\dagger (\tau,0) 
\right) = \rho_{\rm th} + \mathcal{O}(g^2).
\eeq
The above derivation shows that the bath after one kick can be taken to be stationary up to the linear order in the coupling constant $g$. After one kick the 
density matrix of the two-rotor system becomes
\beq
\rho(\tau) = {\rm tr}_B (\rho(\tau)) \otimes \rho_{\rm th} \equiv \rho_S (\tau) \otimes \rho_{\rm th} = \Phi_\tau[\rho_S(0)] \otimes \rho_{\rm th}
\eeq
So, we have shown that in the weak coupling limit, with large bath, the Born 
approximation holds for one kick. Here $\Phi_\tau$ denotes the dynamical map 
from $\rho_S(0)$ to $\rho_S(\tau)$. From the last section, we can conclude 
that $\Phi_\tau$ is a CPTP map as it admits Kraus OSR. Now, for obtaining 
the density matrix after $n$ kicks, we may extend the above definition to
\beq
\rho(n \tau) = {\rm tr}_B (\rho(n \tau)) \otimes \rho_{\rm th} \equiv \rho_S 
(n \tau) \otimes \rho_{\rm th}.
\eeq
However, this straighforward extension using $\hat U(n \tau,0)$ takes us to 
(\ref{non-markov-sys}) and the dynamics remains non-Markovian. To make the 
dynamics Markovian we make one more assumption. We assume that after each 
interaction, the bath quickly resets itself to $\rho_{\rm th}$ and the bath 
at $2 \tau$ is totally uncorrelated to the bath at $\tau$. From a little after
the kick, $\tau^+$, to a little before the next kick, $2 \tau^-$, the system 
and bath evolve separately in their respective Hilbert spaces. We can repeat 
the same procedure for each $\tau$ interval in $n\tau$. Under this, we notice 
that the dynamical map $\Phi_{n\tau}$ can be written as
\beq
\Phi^n \equiv \Phi_{n \tau} = \Phi_\tau \circ \Phi_\tau \circ \cdots \circ 
\Phi_\tau, 
\eeq
where $\circ$ denotes composition of maps. Under this assumption, the dynamical
maps form a discrete Markovian semigroup of CPTP maps 
\beqs
& & \Phi^0 = \mathds{1},\cr
& & \Phi^n \Phi^m = \Phi^m \Phi^n = \Phi^{n+m}.
\eeqs
Under these approximations, the reduced density matrix for the system is 
easily calculated to be
\beq
\rho_S ((n+1) \tau) = \ov{2 \pi} \int_0^{2 \pi} d\th \, e^{-i g \cos(\th - \hat 
\th_S)} e^{-i \pi \hat p_S^2/N} \rho_S(n \tau) e^{i \pi \hat p_S^2/N} 
e^{i g \cos(\th - \hat \th_S)},
\label{CP-2-rotor-map}
\eeq
where $N = 2 \pi m_S/\tau$ is the dimension of the system Hilbert space. For 
small $g$ in the Markovian bath assumption, we observe that the system 
decoheres with the each kick (see Fig.~\ref{fig:CP-map}).

\section{The Lindblad Equation}
\label{GSKL}

Now, we will try to derive the Lindblad equation for the system density matrix of the form
\beq
\dot \rho_S(t) = \mathcal{L} \rho_S(t),
\eeq
where $\mathcal{L}$ is called the Liouvillian superoperator.

We start by going to the interaction picture in which
\beq
\tl A(t) = e^{i(H_S + H_B)t}A e^{-i(H_S + H_B)t},
\eeq
where $A$ represents the operator in the Schr\"odinger picture while $\tl A$ 
represents the operator in the interaction picture. In this picture, the 
von-Neumann equation for time evolution for the system-bath density matrix 
becomes
\beq
\dot{\tl \rho} = -i[\tl H_I,\tl \rho],
\label{von-neumann-eq}
\eeq
where $\tl H_I(t) = e^{i (H_S + H_B)t} \cos(\th_S- \th_B) e^{-i (H_S + H_B)t}$.
By integrating (\ref{von-neumann-eq}) and again substituting the solution in 
this equation, we get
\beq
\dot{\tl \rho}(t) = -i[\tl H_I, \tl \rho(0)] - \int_0^t dt' [\tl H_I(t),[\tl 
H_I(t'), \tl \rho(t')]]
\label{master-eqn-w/o-markov}
\eeq
Now, we trace out the bath degrees of freedom to obtain the time evolution 
equation of the system density matrix. We also use the Born approximation for 
the full density matrix $\rho(t) = \rho_S \otimes \rho_{\rm th}$, where 
$\rho_{\rm th}$ is the thermal density matrix for the bath defined in 
(\ref{bath-simp}). 
Moreover, since $\tr_B\{[\tl H_I, \rho_{\rm th}] \} = 0 $, the above 
simplifications (\ref{master-eqn-w/o-markov}) imply
\beq
\tr_B\{\dot{\tl \rho}(t)\} \equiv \dot{\tl \rho}_S(t) = - \int_0^t dt' \, 
\tr_B\{[\tl H_I(t),[\tl H_I(t'), \tl \rho_S(t') \otimes \rho_{\rm th}]]\}.
\label{master-eqn-to-be-markov}
\eeq
Further, the bath correlations satisy (as shown in the Appendix \ref{decay-bth-corr}),
\beq
\tr_B\{e^{i H_B t'} \cos \th_B e^{i H_B (t-t')} \cos \th_B e^{-i H_B t} \rho_{\rm th} \} = \tr_B\{e^{i H_B t'} \sin \th_B e^{i H_B (t-t')} \sin \th_B e^{-i H_B t} \rho_{\rm th} \} = \fr{1}{2} \del_t^{t'},
\label{correl}
\eeq
as a consequence of which, the system becomes time-local. The Markovian 
assumption is already taken care of by this choice of the bath density matrix. 
Substituting this, the master equation in the Schr\"odinger picture becomes
\beqs
\dot{\rho}_S(t) &=& -i[H_S,\rho_S(t)] + \fr{g^2}{2} \sum_{n=-\infty}^{\infty} \del(t-n \tau)\left( \cos \th_S \rho_S(t) \cos \th_S + \sin \th_S \rho_S(t) \sin \th_S - \ov{2} \{\cos \th_S^2 + \sin \th_S^2, \rho_S\} \right) \cr
&=& -i[H_S,\rho_S(t)] + \fr{g^2}{2} \sum_{n=-\infty}^{\infty} \del(t-n \tau)\left( \cos \th_S \rho_S(t) \cos \th_S + \sin \th_S \rho_S(t) \sin \th_S - \rho_S \right)
\label{lindblad-master-eqn}
\eeqs
where $\{A,B\} = AB + BA$ is the anti-commutator.
Comparing the above equation to the Lindblad master equation
\beq
\dot\rho = -i [H,\rho] + \sum_k \gamma_k (L_k \rho L_k^\dagger - \ov2\{L_k^\dagger L_k , \rho\}),
\eeq
we get the two jump operators $L_1 = \cos\th_S$ and $L_2 = \sin \th_S$. The 
damping rate $\gamma = g^2 \sum_n \del(t-n\tau)/2$ is non-negative. 

Moreover, if we take the bath as a chain of non-interacting identical free 
rotors with an identical coupling to the system rotor, we will have identical 
jump operators $L_k = L_{1,2}$ and damping rates $\gamma_k = \gamma$ for each of the identical free rotors.

The Liouvillian superoperator for (\ref{lindblad-master-eqn}) can be written as
\beq
\mathcal{L}(t) = -i(H_S \otimes \mathds{1}_S - \mathds{1}_S \otimes H_S) + \fr{g^2}{2} \sum_{n=-\infty}^{\infty} \del(t-n \tau) \, (\cos \th_S \otimes \cos \th_S + \sin \th_S \otimes \sin \th_S - \mathds{1}_S \otimes \mathds{1}_S).
\eeq
We notice that the Liouvillian superoperator is $\tau$ periodic $\mathcal{L}(t) = \mathcal{L}(t+\tau)$. Similar to the Floquet operator (time evolution operator for one kick) in the closed system case, we define the flow operator for one kick
\beq
\Lambda^1 \equiv \Lambda_{\tau,0} \equiv \mathcal T \exp \left(\int_0^\tau dt'  \mathcal{L}(t') \right)
 = e^{\fr{g^2}{2}(\cos \th_S \otimes \cos \th_S + \sin \th_S \otimes \sin \th_S - \mathds{1}_S \otimes \mathds{1}_S)} e^{-i \tau(\fr{p_S^2}{2m_S} \otimes \mathds{1}_S - \mathds{1}_S \otimes \fr{p_S^2}{2m_S})},
\eeq

\begin{figure}
    \centering
    \includegraphics[width=0.7\linewidth]{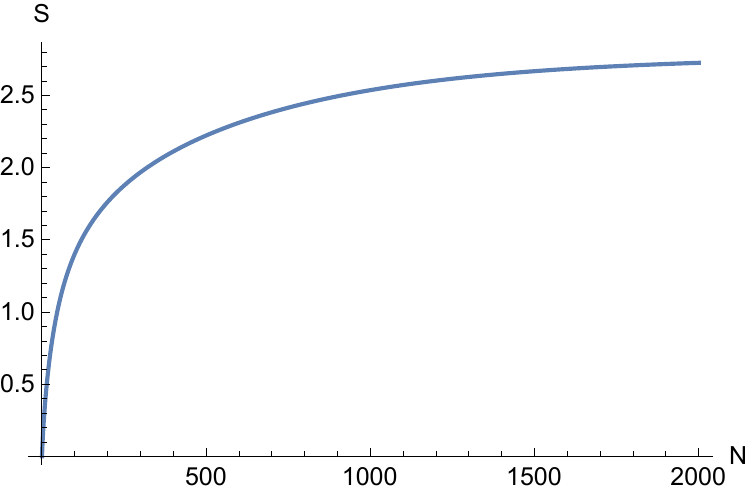}
    \caption{ Plot of von Neumann entropy $S$ vs number of kicks $N$ for the coupling constant $g = 0.1414$. The system starts in the ground state and due to decoherence it approaches the maximally mixed state as $N$ increases.}
    \label{fig:GSKL-evln}
\end{figure}

Here also the dynamical map $\Lambda^n$ forms a discrete Markovian semigroup,
due to the periodic kick. Moreover, we can make the dynamics for open quantum 
system time-independent by making the discrete Markovian semigroup a 
continuous one. For that we consider that the interaction $\hat H_I$ acts at 
every point ($\tau \to 0$) with the kick strength $g \to g' \tau$
\beq
\lim_{\tau \to 0} g' \sum_{n=-\infty}^{\infty} \tau \del(t-n \tau) \approx g' \int_{-\infty}^{\infty} dx \, \del(t - x) = g'.
\eeq
This double scaling limit $\tau \to 0$, $g \to  0 \ni g'$ is fixed, reproduces the two-rotor system under the approximation that the bath Hilbert space is much larger than the system Hilbert space. In this double scaling limit,  we get a Lindblad equation with the time-independent Liouvillian
\beq
\dot{\rho}_S(t) = \mathcal{L} \rho_S (t) = -i[H_S,\rho_S(t)] + \fr{g^2}{2} \left( \cos \th_S \rho_S(t) \cos \th_S + \sin \th_S \rho_S(t) \sin \th_S - \rho_S \right).
\eeq
Note that, for deriving the Lindblad equation for time independent coupling, the correlations in (\ref{correl}) are taken to be Dirac delta $\del(t-t')$, instead of the Kroneker delta for the kicked case. These correlations arrive, when we assume the spectrum of $H_I$ to be continuous instead of discrete. 

\section{Conclusions}
The study of quantum entanglement between two parts of a system, or between 
a system and a suitably defined environment, is a recurring theme in 
understanding various conceptual issues in physics like decoherence, 
diffusion, dissipation, localization, and dynamical chaos. In the present 
paper, we have studied in some detail the entanglement between two quantum 
rotors owing to the coupling between them through a cosine of the angular
distance between them. One of the rotors was integrated out and the von
Neumann entropy of the other rotor was explicitly computed in terms of the
Fourier coefficients that define the Mathieu functions that appear naturally
as eigenfunctions of the Schr\"odinger equation for the simple pendulum 
of the relative coordinate. Introducing a time-periodic delta function 
kick and a series of physically reasonable approximations, allows us to 
derive a master equation of the Lindblad type to describe the time-evolution 
of the reduced system.     

\section{Acknowledgement:} 

This work is partially supported by a grant to CMI from the Infosys Foundation.

\appendix

\section{Decay of Bath Correlations}
\label{decay-bth-corr}

In this appendix we show that the bath correlations decay as $\del_t^{t'}$ (\ref{correl}). To derive this, we use $\cos (\th_S - \th_B) = \cos \th_S \cos \th_B + \sin \th_S \sin \th_B$ and separate the system and bath dependent functions in (\ref{master-eqn-to-be-markov}). Now, we can calculate the bath correlation $\tr_B(e^{i H_B t'} \cos \th_B \, e^{i H_B (t-t')} \cos \th_B \, e^{-i H_B t} \rho_{\rm th})$ by using (\ref{bath-simp}) and putting $H_B = p_B^2/(2 m_B)$. In the basis of energy eigenstates of the bath rotor the above mentioned correlation can be written as
\beqs
&& \tr_B(e^{i H_B t'} \cos \th_B \, e^{i H_B (t-t')} \cos \th_B \, e^{-i H_B t} \rho_{\rm th}) = \ov{2 N_0} \sum_{\nu} \bra \nu | e^{i H_B (t'-t)} \cos \th_B \, e^{-i H_B(t'-t)} \cos \th_B | \nu \ket \cr
= && \ov{2 N_0} \sum_{\nu, \mu} \bra \nu | e^{i H_B (t'-t)} \int \fr{d\th'}{2 \pi} |\th'\ket \bra \th'| \cos \th_B \, e^{-i H_B(t'-t)} |\mu \ket \bra \mu | \int \fr{d\th}{2 \pi} |\th \ket \bra \th | \cos \th_B | \nu \ket \cr
= && \ov{2 N_0} \sum_{\mu, \nu} e^{i \nu^2(t'-t)/(2 m_B)} \int \fr{d\th'}{2 \pi} e^{i (\mu - \nu) \th'} \cos \th' \, e^{-i \mu^2 (t'-t)} \int \fr{d \th}{2 \pi} e^{-i(\mu - \nu) \th} \cos \th \cr
= && \ov{2 N_0} \sum_{\mu, \nu} e^{i (\nu^2 - \mu^2) (t'-t)/(2 m_B)} \ov{2} (\del_\mu^{\nu +1} + \del_\mu^{\nu -1}) \ov{2} (\del_\mu^{\nu -1} + \del_\mu^{\nu + 1}) \cr
= && \ov{2 N_0} \sum_{\mu = -N_0}^{N_0} \ov{2} e^{i (2 \mu + 1) (t-t')/ (2 m_B) }.
\label{correl-val}
\eeqs
So, the correlation term simplifies into a geometric series. The sum of this geometric series for $N_0>>1$, is 
\beq
\lim_{N_0 \to \infty} \ov{4 N_0}  \sum_{\mu = -N_0}^{N_0} e^{i (2 \mu +1) (t- t')/(2 m_B)} = \ov{2} \del_t^{t'}.
\label{correl-val}
\eeq
Similarly, we can calculate $\tr_B(e^{i H_B t'} \sin \th_B \, e^{i H_B (t-t')} \sin \th_B \, e^{-i H_B t} \rho_{\rm th}) = \ov{2} \del_t^{t'}$. Furthermore, by noticing that (\ref{correl-val}) is invariant under the exchange of $t$ and $t'$, we conclude that the correlations are equal.
\beq
\tr_B(e^{i H_B t} \cos \th_B \, e^{-i H_B (t-t')} \cos \th_B \, e^{-i H_B t'} \rho_{\rm th}) = \tr_B(e^{i H_B t} \sin \th_B \, e^{-i H_B (t-t')} \sin \th_B \, e^{-i H_B t'} \rho_{\rm th}) = \ov{2} \del_t^{t'}.
\eeq
Hence, all the bath correlations decay as $\del_t^{t'}/2$.


\begin{thebibliography}{99}

    \bibitem{breur} Breuer, H. P. and Petruccione, F. (2002), {\it The Theory of Open Quantum Systems}, Oxford University Press, Oxford.

    \bibitem{caldiera-leggett} Caldeira, A. O. and Leggett, A. J. (1981), {\it Influence of Dissipation on Quantum Tunneling in Macroscopic Systems}, Physical Review Letters {\bf 46} (4), pp 211–214

    \bibitem{ballentine} Toloui, B. and  Ballentine, L. E. (2009), {\it Quantum localization for two coupled kicked rotors}, arXiv preprint, arXiv:0903.4632.

    \bibitem{chuang} Nielsen, M. A. and Chuang, I. L. (2010), {\it Quantum Computation and Quantum Information: 10th Anniversary Edition}, Cambridge University Press,Cambridge.

    \bibitem{zurek} Zurek, W. H. (1991), {\it Decoherence and the transition from quantum to classical}, Physics Today {\bf 44}(10), pp 36-44. 

    \bibitem{puebla} Puebla, R., Imparato, A., Belenchia, A. and Paternostro, M. (2022), {\it Open quantum rotors: Connecting correlations and physical currents}, Physical Review Research {\bf 4}(4).
    
    \bibitem{arul} Lakshminarayan, A (2001), {\it Entangling power of quantized chaotic systems}, Physical Review E {\bf 64}(3).
    
    \bibitem{santhanam} Santhanam, M. S., Paul, S. and Kannan, J. B. (2022), {\it Quantum kicked rotor and its variants: Chaos, localization and beyond}, Physics Reports {\bf 956}, pp 1-87.

    \bibitem{delande} Delande, D. (2013), {\it Kicked rotor and Anderson localization}, Boulder School on Condensed Matter Physics, Colorado.

    \bibitem{reichl} Reichl, L. E. (2004), {\it The transition to Chaos: Conservative Classical Systems and Quantum Manifestations}, Springer, New York.

    \bibitem{bukov} Bukov, M., D’Alessio, L. and Polkovnikov, A. (2015), {\it Universal high-frequency behavior of periodically driven systems: from dynamical stabilization to Floquet engineering}, Advances in Physics {\bf 64}(2), pp 139–226. 

    \bibitem{mori} Mori, T. (2022), {\it Floquet States in Open Quantum Systems}, Annual Review of Condensed Matter Physics {\bf 14}, pp 35-56.

    \bibitem{buch} Buchleitner, A., Delande, D. and Zakrzewski, J. (2002), {\it Non-dispersive wave packets in periodically driven quantum systems}, Physics Reports {\bf 368}(5), pp 409-547.

    \bibitem{benenti} Rossini, D., Benenti, G. and Casati, G. (2006), {\it Conservative chaotic map as a model of quantum many-body environment}, Physical Review E {\bf 74}(3).

    \bibitem{rivas} Rivas, \'A., Huelga, S. F. and Plenio, M. B. (2014), {\it Quantum non-Markovianity: characterization, quantification and detection}, Reports on Progress in Physics {\bf 77}.

    \bibitem{lindblad} Lindblad, G. (1976), {\it On the generators of quantum dynamical semigroups}, Communications in  Mathematical Physics {\bf 48}(2), pp 119-130.

    \bibitem{matthew-walker} Mathews, J. and Walker, R.L. (1970), {\it Mathematical Methods of Physics. Volume 501}, WA Benjamin, New York.
    
    \bibitem{kraus} Kraus, K. (1983), {\it States, Effects, and Operations}, Lecture
    Notes in Physics {\bf 190} (Springer, Berlin).

    \bibitem{chrusc} Chru\'sci\'nski, D. (2022), {\it Dynamical maps beyond Markovian regime}, Physics Reports {\bf 992}, pp 1-85.


\end{thebibliography}
\end{document}